# Multi-SQL: An extensible multi-model data query language

Yu Yan
Harbin Institute of Technology
yuyan@hit.edu.cn

Nan Jiang
Harbin Institute of Technology
nanjiang1218@gmail.com

Hongzhi Wang
Harbin Institute of Technology
wangzh@hit.edu.cn

YuTong Wang
Harbin Institute of Technology
184009650@qq.com

Chang Liu
Harbin Institute of Technology
liu3980094@163.com

Yuzhuo Wang
Harbin Institute of Technology
wangyzhit@163.com

## ABSTRACT

Big data management aims to establish data hubs that support data in multiple models and types in an all-around way. Thus, the multi-model database system is a promising architecture for building such a multi-model data store. For an integrated data hub, a unified and flexible query language is incredibly necessary. In this paper, an extensible and practical query language–Multi-SQL is proposed to realize the unified management of multi-model data considering the co-processing of multi-model data. To the best of our knowledge, Multi-SQL is the first query language based on various data models. Multi-SQL can also be expanded to suit more complicated scenarios as it is flexible to support other data models. Moreover, we provide a formal semantic definition of the core features of Multi-SQL, including the multi-model definition, multi-model filters, multi-model joins, etc. Furthermore, we propose a two-level query implementation method to totally exploit the existing query optimization capabilities of the underlying engines which could largely improve the query excution efficiency.

## 1. INTRODUCTION

With the advent of the big data age, different types and formats of data have risen explosively. The unified management [1] [2] of multiple data has become a trend, and the demand for a unified and flexible query language has become urgent in databases. Even though existing single-model database management systems have the mature and efficient query languages [3, 4, 5, 6, 7, 8], complex computer application scenarios involve multiple data models, large amount of data and high coupling between data making single-model language inflexible and inefficient. For example, in social commerce[9] which is a promising application, the user information such as user IDs is often stored in a relational table, and the relationships between users are stored as a large graph. When obtaining the relationship between users through IDs, with individual databases for single models, the programmer first needs to query the IDs from the relational database, and then obtain the user relationship from the graph database through IDs, which is pretty complex. From this example, individual single-model databases hardly meet the needs of multi-model data scenarios. Currently, there are two major solutions [10] for multi-model data mangement:(1).unified solution and (2). federated solution.

A unified-model database is an entire database that uses a fully integrated back-end to manage different data models, such as ArangoDB [11], MongoDB [12], PostgreSQL [13] and etc. Those unified solutions are derived from single-model solutions. For example, The JSON-extend query language of MongoDB [12] supports the document, key-value, and graph models, and it is derived from JSON query. The extended SQL in Cassandra could supports relation, key-value, document, and it is developed from SQL. In order to support the query of new data model, unified solutions expand the original query language, such as SQL-extend [14], JSON-extend [15] and etc. Due to using unified query language, the unified approaches which use one basic model to unify multiple models has less challenges in query parsing and query optimization. However, semantic heterogeneity is a very significant challenge in processing multiple datas [16]. Dittrich et al mentioned that "one cannot fit all [17]". The above unified methods have great shortcomings in semantic representation and semantic extension. For example, the JSON-extend query language in MongoDB lacks some important semantic, such as join among relational datas. The CQL in Cassandra is difficult to extend to support graph model because of its simple semantic structure. With the development of modern applications, there will be more widely diverse data models. A mature multi-model query language should have rich semantics and good extensibility.

Table 1: Existing Methods

| Database | Query Language | Supported Data Model | Capable Optimization | Multi-Model Optimization |
|---|---|---|---|---|
| PostgreSQL | extended SQL | Relational, KV, Document | Relational | × |
| IBM DB2 | extended Xquery | relational, Document, Graph | Document | × |
| Cassandra | SQL-like CQL | KV, Document | KV | × |
| DynamoDB | get/put/update | KV, Document, Graph | KV | × |
| MongoDB | extended JSON | KV, Document | Document | × |
| OrientDB | extended SQL | KV, Document, Graph | Graph | × |
| BigDAWG | wrapper, island query | Relational, KV | ALL | cost estimation |
| CloudMdsQL | embedded SQL | Relational, KV, Document, Graph | ALL | cache frequent data |
| Myria | Python API, UDF, UDA | Relational, KV, Graph | ALL | × |
| RHEEM | Middleware | KV, Document, Graph | ALL | × |

In order to efficiently process each type of data, researches studied federated architectures, such as the wrapper of BigDAWG [1], the middleware of RHEEM [2], the embedded SQL CloudMdsQL [18] and etc. The central idea of the federated architecture is to use middlewares, wrappers or embedded solutons to integrate the various query languages. For example, BigDAWG [1] uses the multi-island wrapper to connect multiple data models. CloudMdsQL [18] achieves multi-model query by embedding some underlying stores's APIs to SQL. These federated architectures which consider the features of each data model efficiently retain rich semantics and could totally use the underlying physical query optimization algorithms. However, these methods do not have a unified sematic structures while only implementing a method to integrate various underlying query languages. For example, BigDAWG [1] uses wrappers called shims to connect multiple island while each data island has its own data model and query structure. If there are multiple underlying languages, the federated solutions would lead to difficult multi-model query optimization and low processing efficiency. Semantic heterogeneity has remained a challenge for the federated architecture [16].

The above two kinds of solutions both have shortcomings. Specialized data query languages for their specialized data models are usually more rich semantic and faster processing efficiency than using a unified solution. But with the development of modern applications, federated architectures no longer satisfiy the processing efficiency demand. In this paper, we consider how to support a "specialized service" in a unified language structure for each type of data models. Our Multi-SQL focuses on the following three significant issues:

- Each data model has its own unique semantics, and designing a certain language which could meet the semantic diversity is a big challenge.

- We figure out how to achieve unified coupled query with multiple models which have different and complex structures.

- With the development of modern applications, the expansion of multi-model languages in the new model is an inevitable problem.

Motivated by this, we propose an extensible multi-model query language, Multi-SQL, from a multi-model design perspective which retains the semantic of each data model. To the best of our knowledge, Multi-SQL is the first language which considers multiple data models as its basement. Our language has good scalability and can support the expansion of the new model while using a unified grammatical structure. Our main contributions are as follows:

- We design a multi-model query language, Multi-SQL, and gives the language standard, which solves inflexibility, high cost, and low-efficiency problems of multi-model data management.

- We formalize a unified and diverse data definition language which maintains the abundant semantics of various data models and achieve the unified definition of multiple data models based on a basic unit "TRIPLE".

- We provide a formal specification of data manipulation languages which could flexibly and totally query the multiple data objects, containing atomic filters, multiple selections, complex multiple joins, etc.

- By design two kinds of filters in case of semantic heterogeneity, Multi-SQL could be extended to support new models while do not need to revise the unified grammatical structure.

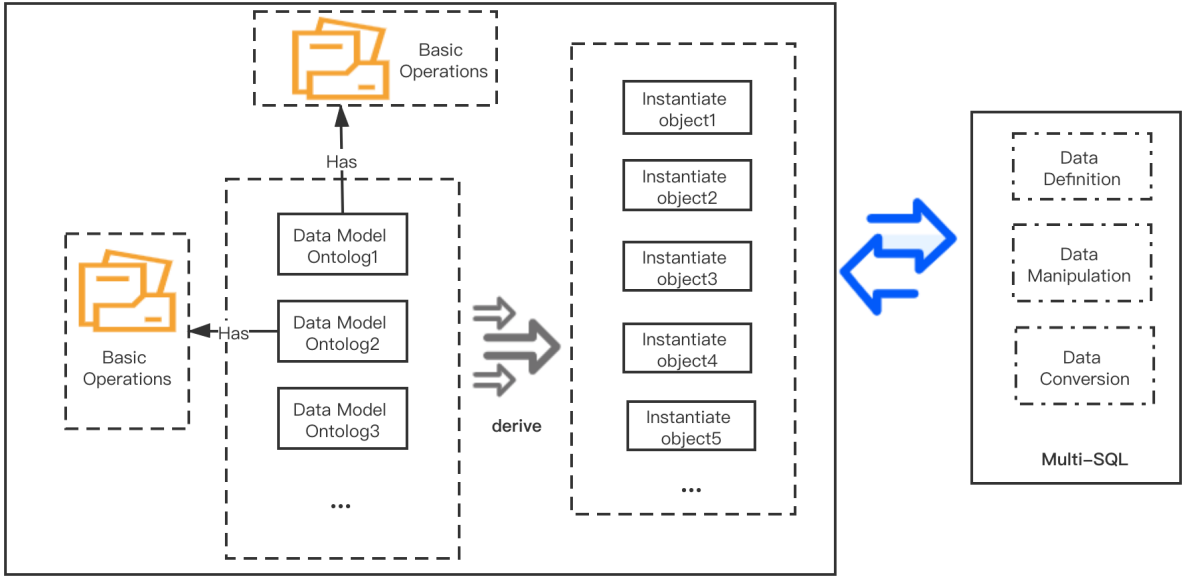

Figure 1: The Overview of Multi-SQL

## 2. RELATED WORK

Existing researches have implemented various query languages in different database systems to handle multiple data models. Initially, SQL [19] was the mainstream query language, which could manage data stored in traditional relational databases. For efficiently processing data, people studied various development for SQL, such as grammar improments, formal definitions and lexical analysis [20, 21, 22], which made SQL more flexible to retrieve, store, modify and delete data. In addition, with the boom of the Internet, XML documents became increasingly popular, so researchers proposed some query syntax for XML data [23, 24, 25]. Moreover, with the increasingly complexity among objects, researchers designed graph data structures including RDF [26], Property Graph, etc. as well as multiple graph query languages including SPARQL, Cypher et al [27, 3, 28, 29, 30].

Furthermore, with the improvement of big data, researchers started to study how to implement complex query across data models. Table 1 shows some popular multi-model query solutions, containing the unified solution and federated architectures. For example, the CQL [31] of Cassandra is a unified solution and could process key-value and JSON. However, to achieve a scalable and flexible architecture in column stores, AQL has many constraints on the syntax, such as the FROM clause can only be followed by one table, the WHERE clause can only be followed by the primary key or the key with a secondary index, etc. These constraints conduct poor efficiency in querying JSON data. The CloudMdsQL [18] is a federated architecture and could process relational data, document and graph. In CloudMdsQL, each data model would be totally processed by the embedded SQL, and it could use all the physical query optimization in underlying store engines. However, if there are multiple underlying languages, the federated solutions would lead to difficult multi-model query optimization and low processing efficiency. And it is difficult for these federated solutions to support mature multi-model optimization, due to its combined structure.

In general, the current multi-modal query methods in the industry have not been designed from the perspective of multiple models, and there are many problems, such as incomplete semantic expression, inefficient query rewriting and inability to reuse related query optimization algorithms. We believe that a unified and extensible query language based on multiple models will be the best way for data hubs in the future.

## 3. OVERVIEW

Multi-SQL is a declarative query language, and it provides the functions for querying and modifying multi-model data and specifying SCHEME definitions. We believe that different data models are interconnected and interdependent in practice, but they are independent in storage. Thus, from a multi-model perspective, each data model needs to be treated equally with independent definitions. At the same time, for flexibility, we design an extensible grammar structure in Multi-SQL. Figure 1 shows the overview of our language. The left

Table 2: Data definition language

```
                              OBJECT CREATION
CREATE <TYPE> <MDNAME>
<TYPE> ::= DOCUMENT | GRAPH | RELATION | KV
                           OBJECT INITIALIZATION
INIT <TYPE> <MDNAME> WITH <OBJECT SCHEME>
<OBJECT SCHEME> ::= <RELATIONAL SCHEME> |<KEY-VALYE SCHEME> | <DOCUMENT SCHEME>
| <GRAPH SCHEME>
<RELATIONAL SCHEME> ::= <TRIPLE>, <TRIPLE> | <RELATIONAL SCHEME>
<KEY-VALUE SCHEME> ::= { <ITEM_NAME, ITEM_TYPE, PRIMARY>, <TRIPLE> }
<DOCUMENT SCHEME> ::= <NESTED TRIPLE>
<GRAPH SCHEME> ::=LIST OF [<NODE> | <EDGE> ]
<NODE> ::= <NODE NAME> { <NESTED TRIPLE>}
<EDGE> ::=<EDGE NAME> { FROM : STRING, TO : STRING, <NESTED TRIPLE>}
<NESTED TRIPLE> ::=<MAP NAME>: <NESTED TRIPLE> | LIST OF <NESTED TRIPLE> | <TRIPLE> |
<NESTED TRIPLE>
<TRIPLE> ::= <ITEM_NAME, ITEM_TYPE, ITEM_CONSTRAINT> | NULL
```
$\text{<VALUE SPACE>} ::= \mathbb{Z} \mid \Sigma \mid TRUE \mid FALSE \mid NULL \mid \text{LIST of <VALUE>} \mid \text{MAP}\{k_1{:}v_m,...,k_m{:}v_m\}$ (m>0, $k_i, v_i \in$ <VALUE>)

```
                               VIEW CREATION
CREATE VIEW <VTYPE> <VIEWNAME> AS | <Q> |
<VTYPE> ::= MULTI | SINGLE
```

box consists of the data model ontologies, the basic operations which is different in each data model and the instantiate objects which are derived from the data model ontologies. For a data model ontology, it could generate many instantiate objects. Each data model ontology has its own basic operations. For example, for the relational model, it could generate many table objects. The relational model has a series operations, such as select, equal, join, etc. Multi-SQL treats these models equally, and constructs the data definition syntax, the data manipulation syntax and the data conversion syntax.

In this paper, we will introduce the syntax and semantic of Multi-SQL in detail. The syntax of Multi-SQL includes three modules, the data definition module, the data manipulation module and the data conversion module. The data definition part is responsible for the definition and initialization of data objects and views. The data manipulation part is responsible for storing, deleting, modifying, and querying data. And the data conversion part charges the data migration between multiple models.

## 4. DATA DEFINITION LANGUAGE

In order to flexibly define multiple data models, we design a unified and diverse data definition language. As shown in Table 2, the keywords of the data definition language mainly include CREATE, RELATION, GRAPH, KV, DOCUMENT, KEY, TRIPLE, etc. For a data model object, we need to create and then initialize it. The initialization is different for each modal. The relation modal needs to specify each column and the primary key of the table. The document model needs to formulate the document pattern. The column names and key-value relationships need to be specified in the KV model. The graph model needs to provide the node structure, edge structure, etc. In this section, we combine the data model formal specification to explain the definition syntax. The data definition languages are specific in Table 2.

Next, we clarify how the data models are defined with some sample examples in Figure 2.

**relational model**.

The relational model are defined as follows in which $A_i$ is the attribute name, and $D_i$ is the domain corresponding to the attribute.

$\mathcal{R}=<(A_0,A_1,A_2...A_n),(D_0,D_1,D_2...D_n)>$
(n=0,1,2...)

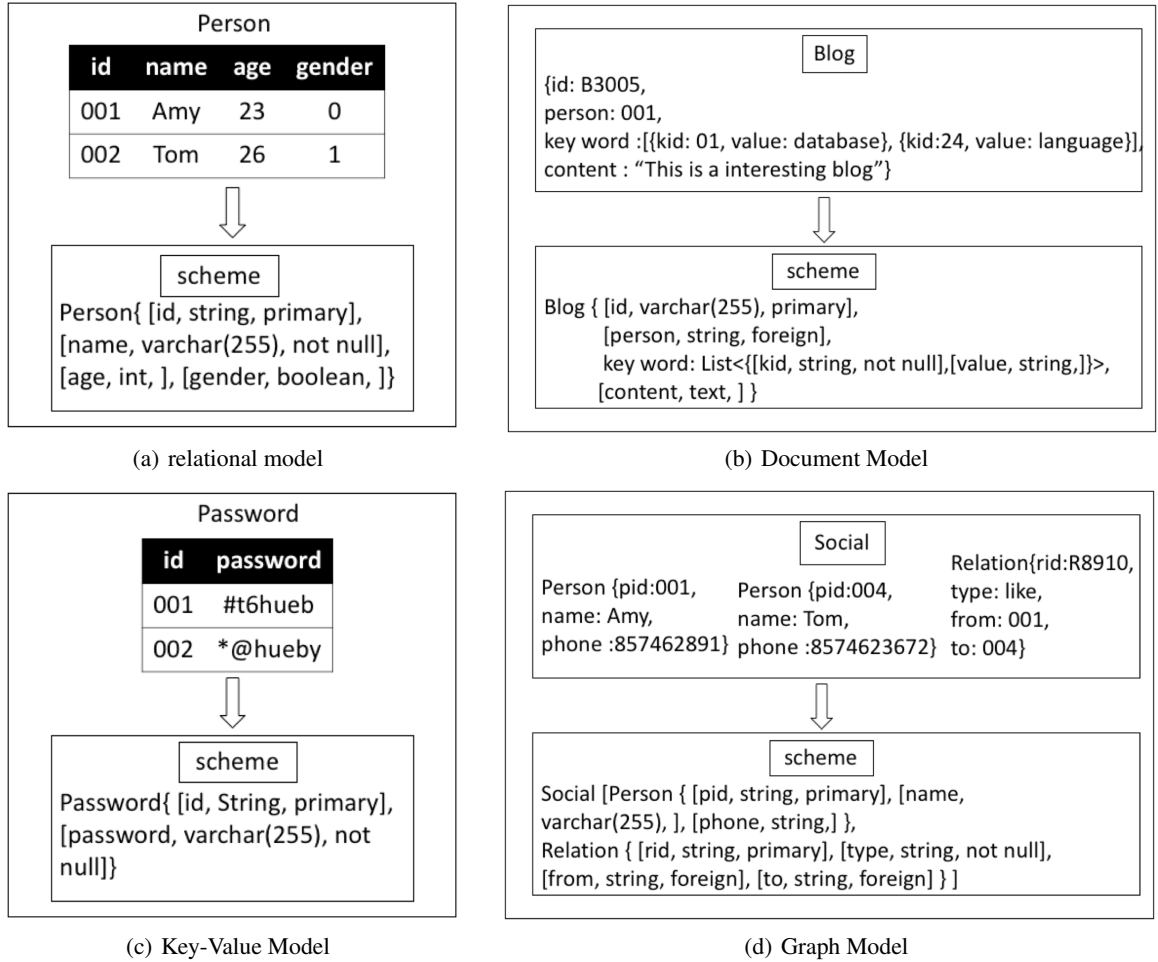

Figure 2: The Examples of Model Definition

As seen in Table 2, a relational table which have four attributes could be modeled as a list of <ITEM_NAME, ITEM_TYPE, ITEM_CONSTRAINT>. In Figure 2(a), the scheme of "Person" table totally expresses the name, data type and constraint of each certain attribute.

**Key-Value Model.**

The key-value structure is defined as follows:
$\mathcal{KV}$ = <(KEY, VALUE), $\zeta$>

For $\forall\ k_i \in$ KEY, $v_i \in$ VALUE, $\zeta(k_i)=v_i$ is a one-to-one mapping. As an example in Figure 2(c), a key-value model consists of a certain primary attribute and a custom attribute. Since key-value is a simplified version of relation, its scheme is similar to the relational model which is a list of the "<TRIPLE>" in Table 2.

**Document Model.**

$\mathcal{D} = (J, ND)$ where $J$ is a free combination of the "<TRIPLE>", and $ND$ is the domain list of document nodes. As the definition in Table 2, the item of "<NESTED TRIPLE>" is either a "<TRIPLE>" or a "<NESTED TRIPLE>". For an example in Figure 2(b), the blog object could be represented by three "<TRIPLE>" and one "<NESTED TRIPLE>" with "key word" as the "<LIST NAME>".

**Graph Model.**

$\mathcal{G} = (N, R, P, T)$, where $N$ is the collection of nodes, $R$ is the collection of edges, $P$ is the collection of properties, and $T$ is the property edges. We formulate the graph as a list of [<NODE> | <EDGE> ] in Table 2. The <NODE> consists of a custom name and some "<NESTED TRIPLE>". The composition of <NODE> has extra two fixed fields (the "<FROM>" and the "<TO>") which specific the head and the tail of an edge. Essentially, we can find that the complex graph model is still a free combination of <TRIPLE>. The example in Figure 2(d) shows a simple social network which has the person node and the relation edge. The scheme of this graph has a <NODE> and a <EDGE>.

In conclusion, all the data models could be expressed by a free combination of attributes which could be specific by a "TRIPLE" in Table 2, <ITEM_NAME, ITEM_TYPE, ITEM_CONSTRAINT>. The difference between data models is reflected in the combination method. As is shown in Section 5, the above models could be directly and efficiently processed by Multi-SQL while existing works are only able to handle a limited number.

**Value Space**.

Moreover, we define the basic "<VALUE SPACE>" which is the foundation of processing data. Using the value space defined in Table 2, multiple data models could be treated. We assume two basic types for further definition: the integers $\mathbb{Z}$, and the type of finite strings over a finite alphabet $\Sigma$. The value set of Multi-SQL is $\mathcal{V}$. Assuming that the set is infinite, the value space is as follows:

(1) The elements of $\mathbb{Z}$ are values.

(2) The elements of $\Sigma$ are values.

(3) True, False and Null are values.

(4) An empty list is a value. If $v_1,...,v_m$ are values, then list$[v_1,...,v_m]$ (m>0) is a value.

(5) An empty map is a value. If $v_1,...,v_m$ are values and $k_1,...,k_m$ are distinct property keys, then the map$\{k_1:v_m,...,k_m:v_m\}$ (m>0) is a value.

The value space of Multi-SQL is abundant, which maintains the diversity of Multi-SQL.

## 5. DATA MANIPULATION LANGUAGE

In this section, we provide a formal specification of the data manipulation language, which is the theoretical basis of Multi-SQL. The formal specification can be used for reasoning to prove the rationality of Multi-SQL, and it is stated in a standardized way that Multi-SQL can support different operations of multiple data models and perform collaborative queries. Besides, when giving the formal specification, we consider the scalability as much as possible to ensure the compatibility for new data models. Data manipulation syntax shown in Table 3, 4 is mainly composed of data query statement, data insertion statement, data update statement, and data deletion statement.

### 5.1 Data Query Language

For each type of language, the query part is the most important component. It is essential to consider both the operation expressions and the efficiency of query optimization in database. Multi-SQL data query statement in Table 3 takes into account the characteristics of each model, and all multi-modal queries is combined by the multiple selections, multiple joins and atomic filters. The above three multiple operations which are in upper level of Multi-SQL are responsible for the combination of the filtering results of the data objects. The atomic filters charge the low-level filtering of the data objects.

#### 5.1.1 Atomic Filter

Due to the different structure of data models, the atomic filters among difference models have some differences. However, these filters still have some common operations in terms of the basic unit("TRIPLE"). We regard these operations in "TRIPLE" which are independent of the data structure and generic as basic filters. In addition to the basic filters, there are also some special filters for each type of data due to their structures, and we call them "characteristic filters". The atomic filters is the foundation of maintaining the semantic diversity in each type of model which has not only some basic operations but also characteristic operations.

We define a finite basic function family $\mathcal{F}$ which is used to implement atomic filters, including some basic filters and characteristic filters.The "<BASIC FILTER>" defined in Table 3 implemented in "TRIPLE" is common and simple. Next, we mainly explain the specific filters of each type of data. By designing the characteristic filters, we increase the diversity of Multi-SQL on the premise of ensuring the uniform.

We mainly consider the four popular data models, containing relation, key-value, document and graph. Importantly, when expanding the new model, users could add special filters to $\mathcal{F}$ for flexibly processing. This change will not affect the semantic parsing because the structure of Multi-SQL is not changed, only the set $\mathcal{F}$ becomes richer.

Firstly, the relational model is a very important model which is the earliest widely-used model. The filters in the relational model contain equal, greater than, less than, etc [20]. In fact, there are a little spacial filters in the relational model and the basic filters in "TRIPLE" could cover most of them.

The most special filter in the relational model is the operation among several corresponding relation instances which is generally called join. And due to the similarity between relational join and multiple join, in Multi-SQL, we unified the connection between several relation instances to the multiple join defined in Section 5.1.2 for being as uniform as possible.

For example, for two relational instances:
$R_1\{(studentid, string, PRIMARY),$
$(name, string,), (class, int,)\},$
$R_2\{(\_id, string, PRIMARY),$
$(studentid, string, FOREIGN), (courseid, string,$
$FOREIGN), (grade, int,)\}$

Table 3: Data Query Language

| QUERY |
|---|
| SELECT <RESULT CONSTRAINT > <RESULT SCHEME> FROM <OBJECT LIST> <WHERE CLAUSE> <ORDER CLAUSE> |
| <RESULT CONSTRAINT > ::= DISTINCT \| NULL |
| <RESULT SCHEME> ::= LIST OF <NESTED ATTRIBUTION> |
| <NESTED ATTRIBUTION> ::= <NESTED ATTRIBUTION> \| LIST OF <NESTED ATTRIBUTION> \| <MAP NAME> : <NESTED ATTRIBUTION> \| <ATTRIBUTION> |
| <ATTRIBUTION> ::= KEY IN <OBJECT> \| NULL |
| <OBJECT> ::= INSTANCE OF <OBJECT SCHEME> |
| <WHERE CLAUSE> ::= WHERE <BASIC FILTER> [AND \| OR \| XOR] <CHARACTERISTIC FILTER> |
| <BASIC FILTER> ::= <BASIC FILTER> [AND \| OR \| NOT \| XOR \| IN \| < \| > \| ≤ \| ≥] <BASIC FILTER> \| <ATTRIBUTION> \| <VALUE> \| NULL |
| <CHARACTERISTIC FILTER> ::= <CHARACTERISTIC FILTER> [AND \| OR \| XOR] <CHARACTERISTIC FILTER> \| <MATCH> \| <PATH> $vert$ NULL |
| [LEFT \| RIGHT] JOIN $O_1, O_2$ RULE <RELATIONAL RULE> WITH <CONNECTED RULE> |
| <RELATIONAL RULE> = A INSTANCE OF <OBJECT SCHEME> <CONNECTED RULE> = <ONE TO ONE> \| <ONE TO MANY> |

we implement a join based on $R_1.studentid = R_2.studentid$ to query the "grade" of "class 3":
SELECT $R_2.\_id, R_1.studentid, R_2.grade$ FROM JOIN $R_1, R_2$
RULE $R_2.\_id, R_1.studentid, R_1.class, R_2.grade$ WITH $R_1.studentid = R_2.studentid$ WHERE $R_1.class = 3$ ORDER BY $R_2.grade$;

For the key-value model, its structure is simple, and it can stably support the storage and efficient query of massive data. The atomic filters could cover all operations in key-value model.

Document data is widely used in the Internet, and its structure is complex. Bourhis.el [15] mentioned that document data has some complex operations due to its recursive structure. The most special operations in document data is the structural query. We take inspiration from the query language in the popular open-source MongoDB, which is also the basis for many other query designs [32]. Multi-SQL regards the complex structural query as a subdocument matching and adds the "MATCH" operations to process them. The result of "MATCH" is a list of documents which satisfy the structured filter condition. MATCH: $\zeta_\tau(\mathbf{D})$, where $\tau$ is a complex structured filter, a nested "attribute_:{expr, value}". The "expr" is one of $\{<, >, \leq, \geq, in\}$ which represents the filter expression. The "value" is one or several values in the domain of the "attribute_". For an example of Figure 2(b), "{id:{=, "BN0024"}},keyword: list<{kid:{in, [01,02]}}>}" is to filter these subdocuments which satisfy id = "BN0024" and "keyword.[∗].kid in [01,02]"

The "MATCH" filter could process all the "<NESTED TRIPLE>", so it could also filter the nodes and edges in graphs. But Francis.et [4] proposed that there are some more complex directional relationship in addition to the linear match in graph model. "MATCH" filter could not process these situations. Inspired by the popular graph query language, we consider to add some directional information to "MATCH". In Multi-SQL, we add "PATH" filter with "→" and "←" coordinating "MATCH" for directional query. PATH: $\Im_\sigma(\mathbf{G})$, where $\sigma$ is a list of "MATCH", "→" and "←". The arrows express the direction of a path. The "MATCH" could filter the nodes and edges. "NODE_MATCH1 → EDGE_MATCH1'→ NODE_MATCH2" expresses simple a path filter with the direction from "NODE_MATCH1" to "NODE_MATCH2".

For an example of Figure 2(d), we could use the blow filter to query all the liked person of Amy's likers: "Person:{name:{=, "Amy"}} → Relation:{type:{=, "like"}} → Person:{} → Relation:{type:{=, "like"}} → Person:{}". First, we use a "MATCH" to filter out the person named Amy. Then we find the persons who have the "like" relationship with Amy. Next, we filter the persons who have a like relationship with Amy's likers. Especially, if we replace a → with ←, the result of the above filter would largely change. For example:"Person:{name:{=, "Amy"}} → Relation:{type:{=, "like"}} → Person:{} ← Relation:{type:{=, "like"}} ← Person:{}". This filter could find the persons who like Amy's likers.

Multi-SQL has good extensibility in the $\mathcal{F}$. If there are some new special operations in a new data model,

Table 4: Other Statements

---

INSERTION

INSERT <OBJECT LIST> MULTIVAL <VAL LIST> | QUERY
<VAL LIST> ::= <VAL LIST>,<VAL LIST> | LIST OF <VALUE> | <VALUE>

UPDATE

UPDATE <OBJECT LIST> SET <VAL LIST> <WHERE CLAUSE>

DELETION

DELETE <OBJECT LIST> <WHERE CLAUSE>

TRANSFER

TRANSFER <O1> INTO <O2> WITH <CO-RELATION>
<CO-RELATION> = LIST OF <ATTRIBUTION> : <ATTRIBUTION>

---

we only need to enrich the $\mathcal{F}$, and do not need to revise the query structure.

### 5.1.2 Multiple Combination

Multiple combinations are in higher level than the atomic filters, containing multiple selections(SELECT)and multiple joins(JOIN). These syntaxs do not change when the product environment brings a new type of data model.

**SELECT.**
$\boxed{\sigma_F[O_1, O_2, O_3, \dots]}$ represents SELECT, which means choosing the subsets from a series multiple objects $[O_1, O_2, O_3, \dots]$ that meet the given conditions $F$. $F$ is the notation of atomic filters, which consists of basic filters and characteristic filters. We take an example to clarify the multiple selection.

SELECT $O_1.a1$ & $O_2.a1, O_2, a2$ &$\{O_3.a1, O_3.a3.name : [O_3.a4, O_3.a5]\}$ WHERE $O_1.a4 = v1$ and $O_2.a8 > v3$ and $O_3.a6 < v4$;

The above multiple selection implements in three objects called $O_1, O_2, O_3$. The $F$ consists of three conditions, containing $O_1.a4 = v1$ and $O_2.a8 > v3$ and $O_3.a6 < v4$. The selection results are three instances of objects, $O_1.a1$ & $O_2.a1$,
$O_2, a2$ &$\{O_3.a1, O_3.a3.name : [O_3.a4, O_3.a5]\}$. In Multi-SQL, the result of a "SELECT" is a list of data objects with "&" as delimiter, and could be processed by "JOIN". It means that any object in selected results could join with other object.

**JOIN.**
$\boxed{(O_1 \bowtie O_2)_\mathcal{R}}$ is an join operation with the join rules $\mathcal{R}$ between two objects. Multiple join is very complex compared to general join. There exists not only one-to-one relationships but also one-to-many relationships. In Multi-SQL, we use JOIN to express the one-to-one join which means that an item in $O_1$ could only be associated with one item in $O_2$. We use the OM JOIN expression the one-to-many join which means that an item in $O_1$ could correspond with many items in $O_2$.

In addition, the main body in the join operation is also important. If we use JOIN, the current two object are both important. LEFT JOIN $\ltimes$ returns all records from the left object, even if there is no matching in the right object. RIGHT JOIN $\rtimes$ and LEFT JOIN $\ltimes$ are the same but in opposite direction.

The join rules, $\mathcal{R}$ consists of two parts, relational rule and connected rule. The relational rule which is a instance of "OBJECT" that explains the relationship between the joined result and $[O_1, O_2]$.

Firstly, we take an example to explain non-nested join:

JOIN $O_1, O_2$ RULE $O_1.a1, O_2.a1, O_2.a2$ WITH $O_1.a1 = O_2.a1$

The result of the join operation is a new relational instance $O_1.a1, O_2.a1, O_2.a2$. JOIN operation could be nested, and could be implemented between all objects and selected results.

For example, a JOIN between a joined result and an object:

JOIN $O_3$, JOIN $O_1, O_2$ RULE $O_1.a1, O_2.a1, O_2.a2$ WITH $O_1.a1 = O_2.a1$ RULE $O_1.a1, O_2.a2, O_3.a4$ WITH $O_3.a1 = O_2.a1$

A JOIN between a selected result and an object:
JOIN SELECT $O_1.a1, O_1.a2$ FROM $O_1$ WHERE $O_1.a1 = v1$, $O_2$ RULE $O_1.a1, O_2.a1, O_2.a2$ WITH $O_1.a1 = O_2.a1$

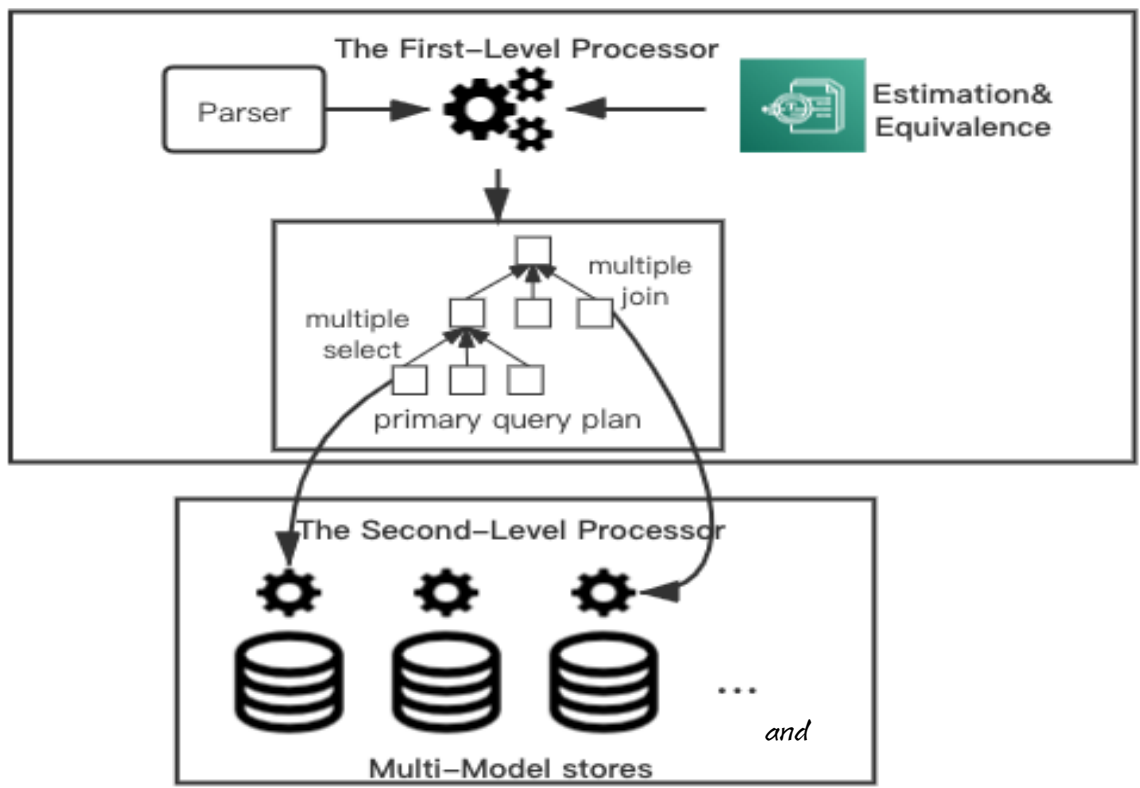

Figure 3: The Status of Multi-SQL

The joined result is very flexible and could be defined by users. Even if there is a demand in the production environment, the joined result of two relational instances can be a document fragment. For example, for two relational instances, $R_1, R_2$, the joined result could be a document: $\{R_1.a1, R_2.a3, R_1.a2.name : \{R_1.a3, R_2.a2\}\}$ Multi-SQL could support the customized design of joined result mainly relying on the flexible definition of <OBJECT SCHEME> which is the combination of <TRIPLE>.

The connected rule shows the connection condition which is the foundation of the join operation. We define two kinds of connected rules. One is <ONE TO ONE> which means that the records of different modalities are in one-to-one correspondence, such as one document record to one relation record. The other is <ONE TO MANY> which means that the records of different modalities are in one-to-many correspondence, such as one relation record to many document record. For example, we join the relational model in Figure 2(a) and the document model in Figure 2(b) to see all the blogs of persons:

JOIN $Person, Blog$ RULE $\{Person.id, blogs : [\{Blog.keyword, Blog.content\}]\}$
WITH $Person.id = Blog.person$

We could find that the joined result is a new document instance and there appears a new attribute "blogs" to serve the key of the "Blog" list. Our Multi-SQL is powerful in support flexible multi-model query.

### 5.2 Other Statements

Since there are little interaction among multiple objects in insert, delete and update compared to the query part, we briefly elaborate on them in Table 4. The data insertion is composed of two parts. <OBJECT> is used to specify the inserted data object. The <VAL LIST> is used to specify the corresponding data and could be the results of certain queries. The data update and data deletion statements also share the similar structure.

With the growth of data volume and the change of computer application workload types, the original data model may be inadequate for the existing data access requirements. We design a model conversion statement that supports the conversion between models, seen in

Table 4. The conversion object can be the data model itself or a query result that meets the constraint specification. The existing mainstream query languages like SQL-extend and CQL have not yet supported data model conversion. Importantly, the conversion correspondence between modes needs to be legal. Duggan.el [33] mentioned some rules of equivalence. For example, it is illegal to convert a large value range D1 to a small value range D2. D1 provides all types floats, from 32 bits to 128bits, while B only supports the float with maximum length of 64 bits. The transformation from D2 to D1 is safe and legal. However, the transformation from D1 to D2 could not be permitted, and it would lost some semantics.

## 6. IMPLEMENTATION

Figure 3 shows the processing of Multi-SQL in a federated system an clarifies the status of Multi-SQL. In our method, a complex multi-model query consists of some atomic filterings and some nested join operations. After parsing the multi-model query, the system needs the first-level process to optimize the multiple filter, join, etc. based on the cardinality estimations and semantic equivalences. The system would get a series of atomic operations and deliver them to the store engines. The second-level process is to optimize the certain subquery based on a certain store, which could use almost all the existing query optimal algorithms based on I/O cost, node connection cost, indexes, etc. Similar to Garlic [34], we advocate a query optimal algorithm which could exploits the existing query optimization capabilities of the underlying engines. Considering the existing physical query optimization algorithms and multi-model query optimization, Multi-SQL could be implemented by a separated two-layer language structure so that most of existing optimal algorithms could directly work. The open-source parser and some examples is avaiable at github[1].

## 7. CONCLUSION

In this paper, we achieve the first attempt to propose a formal multi-model query language. The multi-model database is increasingly prevalent across a wide variety of application domains. There is a requirement for the query language to allow for multi-model operations to be expressed directly. As the first multi-model query language designed from a multi-model perspective, Multi-SQL is a solidly-established declarative query language with scalability, flexibility, and efficiency. All the work, including the multi-model definition semantics and complex query semantics described in this paper, will help establish a data hub which fully supports data of various models and types. Apart from these fundamental designs in our paper, we will pay more attention to the implement of Multi-SQL in Datahubs which is a very significant practical problems in the future. In particular, we believe that this area is worthy of further research.

---

[1] https://github.com/yy19970618/Multi-SQL/tree/master